\documentclass[a4paper,11pt]{article}
\usepackage[a4paper,left=0.99in, right=0.99in,top=1.2in, bottom=1.2in]{geometry}

\usepackage{hyperref}
\usepackage{url}
\usepackage{amsmath}
\usepackage[dvips]{graphicx}
\usepackage[dvipsnames]{xcolor}
\usepackage{cite}

\newcommand{\TMD}{TMD}
\newcommand{\TMDs}{TMDs}
\newcommand{\gluonTMD}{gluon\ TMD}
\newcommand{\gluonTMDs}{gluon\ TMDs}

\title{Sudakov effects in central-forward dijet production\\ in high energy factorization}

\author{
A.~van Hameren,$^1$ P.~Kotko,$^2$ K.~Kutak$^1$ and S.~Sapeta$^1$\\\\
$^1$ 
{\small\it The H.\ Niewodnicza\'nski Institute of Nuclear Physics PAN,}\\ 
{\small\it Radzikowskiego 152, 31-342 Krak\'ow, Poland}\\\\
$^2$ 
{\small\it AGH University of Science and Technology, Physics Faculty,}\\
{\small\it Mickiewicza 30, 30-059 Krak\'ow, Poland}\\
}

\date{}
\begin{document}
\maketitle

%--------------- preprint numbers ---------------------
\vspace{-25em}
\begin{flushright}
  IFJPAN-IV-2020-8
\end{flushright}
\vspace{20em}
%-----------------------------------------------------------------

\begin{abstract}
We discuss central-forward dijet production at LHC energies within the
framework of high energy factorization.
In our study, we profit from the recent progress on consistent merging of
Sudakov resummation with small-$x$ effects, which allows us to compute two
different gluon distributions which depend on longitudinal momentum, transverse
momentum and the hard scale of the process: one for the quark channel and one
for the gluon channel.
The small-$x$ resummation is included by means of the BK equation
supplemented with a kinematic constraint and subleading corrections.
We test the new gluon distributions against existing CMS data for transverse
momentum spectra in forward-central dijet production.  We obtain results which
are largely consistent with our earlier predictions based on model
implementation of Sudakov form factors. In addition, we study dijet azimuthal
decorrelations for the forward-central jets, which are known to be sensitive to
the modeling of soft radiation.
\end{abstract}

%-----------------------------------------------------------------------------
\section{Introduction}
\label{sec:introduction}

Processes with jets remain one of the most important tools used to study Quantum
Chromodynamics (QCD) at hadron colliders, in particular at the
LHC~\cite{Salam:2009jx, Sapeta:2015gee} and future Electron Ion
Collider~(EIC)~\cite{Accardi:2012qut,Li:2020rqj,Arratia:2019vju,Zheng:2014vka}.
Amongst them, production of dijets proves particularly useful to address various
questions concerning QCD dynamics.  When both jets are produced in the central
rapidity region, the energy fractions of the incoming partons are comparable and
sizable.  Theoretical predictions for such configuration can be safely
calculated in the framework of collinear factorization. However, when one of the
jets moves in the forward direction, $y_\text{jet} \gg 0$, one of the incoming
hadrons is probed at relatively low momentum fraction $x$, and that leads to the
appearance of large logarithms $\ln x$, which have to be resummed. The optimal
description of this process is achieved within the hybrid
factorization~\cite{Catani:1990eg,Dumitru:2005gt,Marquet:2007vb,Deak:2009xt},
where the matrix elements are evaluated with one of the incoming partons being
off-shell.  The momentum distribution of that parton obeys the BFKL equation
\cite{Balitsky:1978ic,Kuraev:1976ge,Fadin:1975cb,Kovchegov:2012mbw}, which
depends not only on the longitudinal part of the momentum, but also on its
transverse component. We will from now on refer to these as
\emph{transverse momentum dependent} parton distributions (TMDs).
In addition, when both jets move forward, the value of~$x$ is even smaller and
one starts being sensitive to saturation effects~\cite{Gribov:1984tu,
Mueller:1985wy}. The corresponding evolution equation becomes
nonlinear~\cite{Balitsky:1995ub,
Kovchegov:1999yj,JalilianMarian:1997jx,JalilianMarian:1997gr,JalilianMarian:1997dw},
as density of gluons at low~$x$ is  very high.

While the small-$x$ effects can be taken into account by using one of the
phenomenologically successful \TMDs, there is another class of effects relevant
for forward jet production which should also be accounted for, namely the
resummation of Sudakov logarithms. They are important as the hard scale provided
by jet transverse momentum opens phase space for logarithmically enhanced soft
and collinear emissions
\cite{Dokshitzer1980,Lipatov:2019nbx,Collins1985,Bury:2017jxo,VanHaevermaet:2020rro}.
See also recent Monte Carlo developments where one constructs TMD distributions
that account for $k_T$ and Sudakov effects \cite{Hautmann:2017fcj}.

As demonstrated in Refs.~\cite{Mueller:2013wwa,Mueller:2012uf, Sun:2015doa,
Sun:2014gfa, Mueller:2016xoc}, small-$x$ and Sudakov resummations can be
performed simultaneously in $b_\perp$ space and can then be cast into transverse
momentum dependent gluon distributions.  Such \TMDs\ have already been used in
phenomenological calculations of di-hadron correlations at
EIC~\cite{Zheng:2014vka} and in proton-nucleus collisions at RHIC
\cite{Stasto:2018rci,Marquet:2019ltn}. While in
\cite{Zheng:2014vka,Stasto:2018rci}  the Golec-Biernat-W\"usthof
model~\cite{GolecBiernat:1998js} was employed to account for small-$x$ effects,
in \cite{Marquet:2019ltn} the rcBK was used.  

In the present work, we focus on Sudakov effects in  the process of
central-forward dijet production in proton-proton collisions.
Similarly to previous studies
\cite{Deak:2010gk,Deak:2011ga,Kutak:2012rf}, we perform our calculations in the
framework of \emph{high energy factorization} (HEF) factorization
\cite{Gribov:1984tu, Catani:1990eg, Catani:1994sq, Collins:1991ty}, where the
cross section is calculated as a convolution of a hard
sub-process~\cite{Lipatov:1995pn,Antonov:2004hh} and nonperturbative parton
densities, which take into account longitudinal and transverse degrees of
freedom. At low $x$, gluons dominate over quarks, hence we consider only gluon
\TMDs.

In our earlier study of the central-forward dijet
production~\cite{vanHameren:2014ala}, the Sudakov effects were introduced by
means of a simplified procedure~\cite{vanHameren:2014ala, Kutak:2014wga}, which
nevertheless turned out to be phenomenologically successful.  We then used the
same approach to study forward-forward dijet production in proton-proton and
proton-lead collisions, focusing on the broadening of the dijet azimuthal
correlation spectrum  \cite{vanHameren:2019ysa}. We found that a delicate
interplay between the Sudakov effects and the saturation effects is needed to
describe the LHC data.  Although our phenomenological Sudakov model works well
in that regard, a more systematic calculation of the Sudakov effects is
necessary in order to solidify the predictions. One of the difficulties here
comes from the proliferation of the small-$x$ TMD gluon distributions needed in
the saturation regime \cite{Dominguez:2011wm,Kotko:2015ura}. The Sudakov
resummation affects all these distributions in a rather complicated way. The
following work is a first step towards a fully general approach and focuses on a
single small-$x$ TMD gluon distribution, which appears in inclusive processes
and in situations where saturation effects are mild.

In this work, we use the proper Sudakov factors derived within perturbative
QCD~\cite{Mueller:2013wwa, Mueller:2012uf, Sun:2014gfa, Sun:2015doa,
Mueller:2016xoc} and profit from the recent progress on consistent merging of
Sudakov resummation with small-$x$ effects~\cite{Stasto:2018rci}.  These new
elements allow us to significantly elevate theoretical status of our predictions
for the discussed process of interest.
We shall then compare the upgraded results to the ones which used simplistic
models of including the Sudakov effects into the small-$x$ gluon, as well as to
available experimental data. 
As in Ref.~\cite{vanHameren:2014ala}, the modeling of the small-$x$ effects in
this work comes from the BK equation supplemented with a kinematic constraint
and subleading corrections~\cite{Kutak:2012rf}.

For the central-forward configuration of the final-state jets, one of the
longitudinal fractions of the hadron momenta is much smaller then the other,
$x_{B}\ll x_{A}$. This follows from simple kinematic relations
\begin{equation}
  x_A = \frac{1}{\sqrt{s}}\left(|p_{1\perp}|  e^{y_1} + 
                                 |p_{2\perp}| e^{y_2}\right)\,,
  \qquad
  x_B = \frac{1}{\sqrt{s}}\left(|p_{1\perp}| e^{-y_1} + 
                                |p_{2\perp}| e^{-y_2}\right)\,,
  \label{eq:xAxB}
\end{equation}
where $\sqrt{s}$ is the center-of-mass energy of the proton-proton collision,
while $p_{i\perp}$ and $y_{i}$ are the transverse momenta (Euclidean
two-vectors) and rapidities of the produced jets.
The formula for the \emph{hybrid} high energy factorization
reads~\cite{Dumitru:2005gt,Deak:2009xt}
\begin{align}
  d\sigma_{A+B\rightarrow j_1 + j_2 + X} & = 
  \int dx_{A}\, 
  \int\frac{dx_{B}}{x_{B}}\,
  \int \frac{d^{2}k_{B\perp}}{\pi}
  \nonumber \\[0.5em]
  &
  \hspace{10pt}
  \times
  \sum_{a, c, d}
  f_{a/A}\left(x_{A},\mu \right)\,
  \mathcal{F}_{g^{*}/B}\left(x_{B},k_{B\perp},\mu \right)\,
  d\hat{\sigma}_{a+g^{*}\rightarrow c+d}
  \left(x_{A},x_{B},k_{B\perp},\mu \right),
  \label{eq:hef}
\end{align}
where $\mathcal{F}_{g^{*}/B}$ is the so-called  \emph{unintegrated gluon
density} or \emph{transverse momentum dependent gluon distribution} (see
\cite{Kharzeev:2003wz,Dominguez:2011wm,Kotko:2015ura,Bury:2018kvg} for more
details on different gluon distributions), $f_{a/A}$ are the collinear PDFs and
$d\hat{\sigma}_{a+g^{*}\rightarrow c+d}$ is built out of the off-shell
gauge-invariant matrix elements. The indices $a, c, d$ run over the gluon and
all the quarks that can contribute to the inclusive dijet production. 
Notice that both $f_{a/A}$ and $\mathcal{F}_{g^{*}/B}$ depend on the hard
scale~$\mu$, and the latter depends also on the transverse momentum of the
incoming gluon, whose value is linked to the final-state kinematics by the
relation
\begin{equation}
  |k_\perp|^2 = |p_{1\perp}+p_{2\perp}|^2 = 
  |p_{1\perp}|^2+|p_{2\perp}|^2+ 2 |p_{1\perp}| |p_{2\perp}| \cos\Delta\phi\,,
  %\label{eq:}
\end{equation}
where $\Delta\phi$ is the azimuthal distance between the jets.
The hard scale dependence in the \TMD\ is necessary to properly account for
large Sudakov logarithms that appear predominantly in the back-to-back region,
where $k_{\perp}$ is small, but $\mu$ remains large for relatively hard jets.
As shown in Ref.~\cite{vanHameren:2014ala}, incorporating the hard scale
dependence in the \TMD\ is essential to successfully describe shapes of dijet
spectra.

It is important to mention that, as discussed in Ref.~\cite{Kotko:2015ura}, the
high energy factorization formula~(\ref{eq:hef}) is valid only when $Q_s \ll
|k_\perp|  \ll |p_{1\perp}|, |p_{2\perp}|$, which corresponds to collisions of
relatively dilute hadrons. The process of central-forward dijet production in
$p-p$ collision, which is the focus of our study, corresponds exactly to that
situation.  For processes which involve dense targets, like for example
forward-forward dijet production in $p-A$ collisions, Eq.~(\ref{eq:hef}) has to
be replaced by a more general factorization formula with multiple transverse
momentum dependent gluon distributions~\cite{Dominguez:2011wm,
Kotko:2015ura,vanHameren:2016ftb,Altinoluk:2019fui}.

%-----------------------------------------------------------------------------
\section{Dipole gluon with Sudakov form factor}
\label{sec:sudakov}

The Sudakov effects are most conveniently included in position space.
The resulting \gluonTMD, which incorporates both small-$x$ and soft-collinear
resummation, can be then transformed to momentum space as
follows~\cite{Stasto:2018rci}
\begin{equation}
  \mathcal{F}_{g^*/B}^{ag\to cd}(x, q_\perp, \mu) = \frac{-N_c S_\perp}{2\pi
  \alpha_s}
  \int_0^{\infty} \frac{b_\perp db_\perp}{2\pi} J_0(q_\perp b_\perp)\,
  e^{-S_{\textrm{Sud}}^{ag\to cd}(\mu, b_\perp)}\, \nabla_{b_\perp}^2
  S(x,b_\perp)\,,
  \label{eq:Fqgpos}
\end{equation}
where $S_{\perp}$ is the transverse area of the target and $S(x,b_{\perp})$ is
the so-called dipole scattering amplitude, which in the Color Glass Condensate
(CGC) theory (see {\it e.g.} \cite{Gelis:2010nm}) is related to the color
average of the dipole operator, {\it i.e.} two infinite Wilson lines displaced
in the transverse plane.
(Notice the difference in the prefactor w.r.t.\ to Ref.~\cite{Stasto:2018rci},
which comes from the fact that $\mathcal{F}_{^{g^*/B}} = \pi {\cal F}_{^{qg}}^{_{(a)}}$.)
The Sudakov factors come from the resummation of soft-collinear gluon radiation
and they depend on the partonic channel. Hence, the gluon with the Sudakov
acquires this dependence and, consequently, a single dipole gluon is replaced
with a set of gluons $\{\mathcal{F}_{^{g^*/B}}^{ab\to cd}\}$. In practice, the
two channels that dominate in the central-forward productions are: $qg \to qg$
and $gg \to gg$. Hence, we will need to determine two \gluonTMDs:\
$\mathcal{F}_{^{g^*/B}}^{qg\to qg}$ and $\mathcal{F}_{^{g^*/B}}^{gg\to gg}$.

It is appropriate to mention that in our study we resort to the so-called
mean-field approximation (known to work very well, see {\it e.g.}
\cite{Dumitru:2011vk}), which allows one to calculate quadrupole operators in
terms of the dipole operators alone and thus to use the BK equation for
evolution of the gluon density.

By taking the Fourier transform of Eq.~(\ref{eq:Fqgpos}), we can express the
gluon with Sudakov resummation by the gluon without the Sudakov, all in momentum
space
\begin{equation}
    {\cal F}_{g^*/B}^{ab\to cd}(x,k_\perp,\mu) =
    \int d b_\perp \int dk^{\prime}_\perp \, b_\perp\, k^\prime_\perp\, 
    J_0(b_\perp\,k^\prime_\perp)\, 
    J_0(b_\perp \,k_\perp)\,
    {\cal F}_{g^*/B}(x,k^\prime_\perp)\, e^{-S_\text{Sud}^{ab\to cd}(\mu,b_\perp)}\,.
    \label{eq:gluon-dff}
\end{equation}
For each channel, the Sudakov factors can be written as
\begin{equation}
  S_{\text{Sud}}^{ab\to cd} (b_\perp) =\sum_{i=a,b,c,d} S_p^{i} (b_\perp) +
\sum_{i=a, c, d}S^{i}_{np} (b_\perp), 
\end{equation} 
where $S_p^{i} (b_\perp) $ and $S_{np}^{i} (b_\perp)$ are the perturbative and
non-perturbative contributions. 
As argued in~Ref.~\cite{Stasto:2018rci}, as small-$x$ \gluonTMDs\ for parton $b$
may already contain some non-perturbative information at low-$x$, the
non-perturbative Sudakov factor associated with that incoming gluon~$b$ should
not be included. 
In addition, according to the derivation in Ref.~\cite{Mueller:2013wwa}, the
single logarithmic term in the perturbative part of the Sudakov factor -- the
so-called $B$-term --  should also be absent for the incoming small-$x$ gluon. 
The perturbative Sudakov factors are given by~\cite{Stasto:2018rci}
\begin{eqnarray}
S_p^{qg\to qg} (Q, b_\perp) & = & \int_{\mu_b^2}^{Q^2} \frac{d\mu^2}{\mu^2} \left[
2 (C_F + C_A) \frac{\alpha_s}{2\pi} \ln \left( \frac{Q^2}{\mu^2} \right)
- \left(\frac{3}{2}C_F +  C_A \beta_0 \right) \frac{\alpha_s}{\pi}
  \right],
  \label{eq:sudpertqg}
  \\[1em]
S_p^{gg\to gg} (Q, b_\perp) & = & \int_{\mu_b^2}^{Q^2} \frac{d\mu^2}{\mu^2} \left[
  4 C_A \frac{\alpha_s}{2\pi} \ln \left( \frac{Q^2}{\mu^2} \right)
- 3 C_A \beta_0 \frac{\alpha_s}{\pi} \right]\,,
  \label{eq:sudpertgg}
\end{eqnarray}
where 
$\beta_0 = (11-2n_f/3)/12$, $\mu_b=2e^{-\gamma_E}/b_*$, and $b_* =
b_\perp/\sqrt{1+b_\perp^2/b_{\rm max}^2}$. 
The $g g \to q \bar q$ channel is negligible for the kinematics of this
study.
Following Ref~.\cite{Stasto:2018rci}, for the non-perturbative Sudakov
factor, we employ the parameterization~\cite{Su:2014wpa,Prokudin:2015ysa}
\begin{eqnarray}
\label{eq:sudnpqg}
S_{np}^{qg\to qg} (Q, b_\perp) & = & 
\left( 2 + \frac{C_A}{C_F} \right) \frac{g_1}{2} b_\perp^2 + \left(2 +
\frac{C_A}{C_F} \right) \frac{g_2}{2} \ln \frac{Q}{Q_0} \ln
\frac{b_\perp}{b_*}\,, \\
\label{eq:sudnpgg}
S_{np}^{gg\to gg} (Q, b_\perp) & = & 
\frac{3C_A}{C_F} \frac{g_1}{2} b_\perp^2 + \frac{3C_A}{C_F} \frac{g_2}{2} \ln
\frac{Q}{Q_0} \ln \frac{b_\perp}{b_*}\,,
\end{eqnarray}
with $g_1 = 0.212$, $g_2 = 0.84$, and $Q_0^2 = 2.4 {\rm GeV}^2$. 

\begin{figure}[p]
  \begin{center}
    \includegraphics[width=0.99\textwidth]{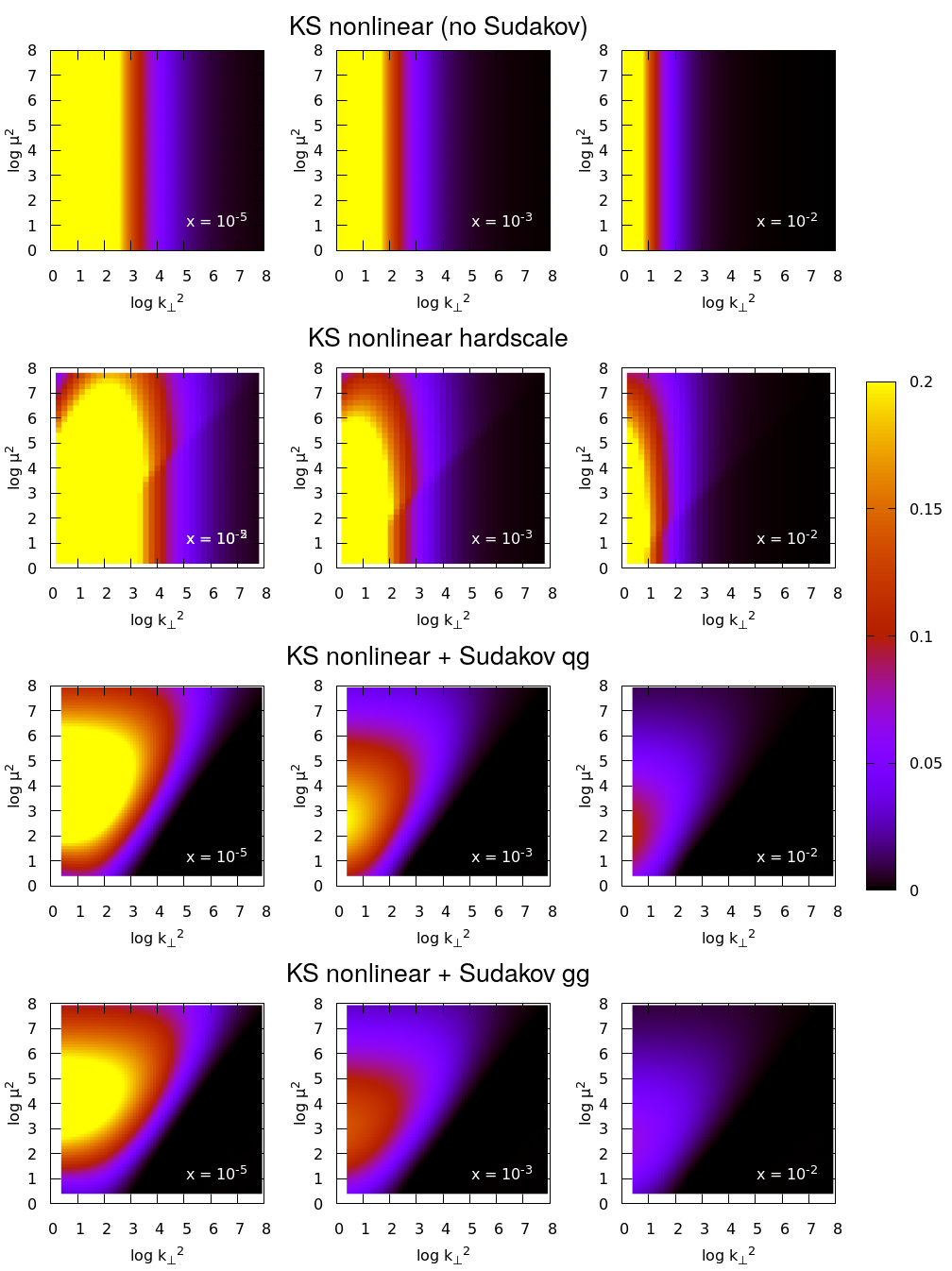}
  \end{center}
  \caption{
  KS gluon distribution, Eq.~(\ref{eq:gluon-dff}) without and with the Sudakov
  form factors. The second row corresponds to the simple model-Sudakov given in
  Eq.~(\ref{eq:survival}), while the third and the fourth rows show results
  obtained with the Sudakov factors derived from QCD and given in
  Eqs.~(\ref{eq:sudpertqg}), (\ref{eq:sudnpqg}), and (\ref{eq:sudpertgg}), 
  (\ref{eq:sudnpgg}), respectively.
  }
  \label{fig:gluon-map}
\end{figure}

As a basis for all calculations presented in this study, we use the nonlinear KS
(Kutak-Sapeta) \gluonTMD~\cite{Kutak:2012rf}, which, for $k_\perp^2>1\,
\text{GeV}^2$, comes from evolving the input distribution
\begin{equation}
 \mathcal{F}_{{g^*/B}}^{(0)}(x,k^2_\perp)=
 \frac{\alpha_S(k^2_\perp)}{2\pi k^2_\perp}\int_x^1
 dzP_{gg}(z)\frac{x}{z}g\left(\frac{x}{z}\right)\,,
 \quad
 \text{where}
 \quad
 xg(x)=N(1-x)^{\beta} (1-D x)\,,
 \label{eq:initial-cond} 
\end{equation}
with the extension of the BK (Balitsky-Kovchegov) equation \cite{Kutak:2003bd},
following the prescription of Ref.~\cite{Kwiecinski:1997ee} 
to include kinematic
constraint on the gluons in the chain, non-singular pieces of the splitting
functions, as well as contributions from sea quarks. 
For $k_\perp^2 \leq 1\, \text{GeV}^2$, the gluon distribution is taken as
$\mathcal{F}_{^{g^*/B}} (x,k^2_\perp) = k^2_\perp
\mathcal{F}_{^{g^*/B}}(x,1)$, which is motivated by the shape obtained
from the solution of the LO BK equation in the saturation regime
\cite{Sergey:2008wk}.

The parameters of the gluon were set by a fit to the $F_2$ data from
HERA~\cite{Aaron:2009aa}, which returned the values: $N=0.994$, $\beta= 18.6$,
$D=-82.1$ and $R=2.40\, \text{GeV}^{-1}$.  The first three parameters correspond
to the initial condition given in Eq.~(\ref{eq:initial-cond}), while the last
parameter is responsible for the strength of nonlinear effects in the evolution
equation.  The overall quality of the fit was good, with $\chi^2/\text{ndof} =
1.73$.

We emphasise that the gluon constrained by the above fit can be used in our
study without any modifications. This comes from the fact that it corresponds
to the small-$x$ kinematic regime and it is universal amongst DIS and
central-forward jet production processes~\cite{Kotko:2015ura}, where saturation
effects are moderate. 
The same is true for the Sudakov
factors used in our study. The perturbative part is parameter-free while the
non-perturbative terms are universal in the kinematic domain of our
study~\cite{Su:2014wpa}. Moreover due to the high transverse momenta of the
final state jets, non-perturbative effects in the Sudakov are less important
than in the case of hadron production.

We introduce the Sudakov effects into the KS gluon distribution following the
formalism described above. In addition, for reference, we use two methods
employed in our earlier studies~\cite{vanHameren:2014ala, Kutak:2014wga}.  Those
calculations used the Sudakov form factor, understood as the DGLAP evolution
kernel, that has been applied on the top of the \gluonTMD, together with
constrains such as unitarity. Those methods should therefore be considered as
models, in contrast to the proper resummation of Sudakov logarithms considered
in this work.
Nevertheless, the approaches used in Refs.~\cite{vanHameren:2014ala,
Kutak:2014wga} were phenomenologically successful (see
also~\cite{vanHameren:2019ysa}), and one of the objectives of this study is to
check how the predictions of those simplistic models compare with the proper way
of including the Sudakov effects into the small-$x$ gluon. The reference models
are:
\begin{itemize}
  %\justifying
  %\setlength\itemsep{1.2em}
  \item
  Model 1:
  The survival probability model~\cite{vanHameren:2014ala}, where the Sudakov
  factor of the form \cite{Watt:2003mx}
  \begin{equation}
    T_s(\mu_{F}^2,k_{\perp}^2)=
    \exp\left(-\int_{k_{\perp}^2}^{\mu_F^2}\frac{dk_{\perp}^{\prime
    2}}{k_{\perp}^{\prime 2}}\frac{\alpha_s(k^{\prime 2}_{\perp})}
    {2\pi}\sum_{a^\prime}\int_0^{1-\Delta}dz^{\prime}P_{a^\prime
    a}(z^\prime)\right)\,,
    \label{eq:survival}
  \end{equation}
  is imposed at the level of the cross section.  This procedure
  corresponds to performing a DGLAP-type evolution from the scale $\mu_0\sim
  |\vec{k}_{\perp}|$ to $\mu$, decoupled from the small-$x$ evolution. 

  \item
  Model 2:
  The model with a hard scale introduced in Ref.~\cite{Kutak:2014wga}. The
  Sudakov form factor of the same form as in Eq.~(\ref{eq:survival}) is imposed
  on top of the KS gluon distribution in such a way that, after integration of
  the resulting hard scale dependent \gluonTMD, one obtains the same result as
  by integrating the KS gluon distribution.
\end{itemize}

In Fig.~\ref{fig:gluon-map} we show the KS gluon distributions, with and without Sudakov form
factors, as functions of the transverse momentum $k_\perp$ and the hard scale
$\mu$. Three columns correspond to three different $x$ values. The first row
shows the original KS gluon distribution, which, as expected, does not depend on the value of
$\mu$. In the second row, we show the KS hardscale gluon distribution of
Ref.~\cite{Kutak:2014wga} (the other  model~\cite{vanHameren:2014ala} does
not allow one to plot gluon distribution, as it applies Sudakov effects
at the cross section level via a reweighting procedure).
Here, the dependence on $\mu$ is non-trivial and we see that the gluon develops
a maximum in that variable. As shown in the figure, this maximum is rather
broad.  In the third and the fourth row of Fig.~\ref{fig:gluon-map}, we present
our new KS gluon distribution with the Sudakov form factor described in this section. As
explained earlier, this gluon exists in two versions, one for the $qg$ and the
other for the $gg$ channel. The dependence on $k_\perp$ and $\mu$ is
qualitatively similar between the new gluons and the naive KS hardscale gluon
distribution.
In the former case, however, the peak is significantly narrower in $\mu$ as
compare to the naive model of Ref.~\cite{Kutak:2014wga}. It is interesting to
note that the $qg$ gluon is broader than the $gg$ gluon. This can be understood
by comparing the colour factors in the Sudakov functions~(\ref{eq:sudpertqg})
and (\ref{eq:sudpertgg}). The colour factor is bigger for the $gg$ channel,
hence, in that case, the Sudakov suppression is stronger along the $\mu$
direction.

We have as well computed linear versions of the KS gluon distributions with the Sudakov,
using the KS linear gluon distribution of Ref.~\cite{Kutak:2012rf}. We also used them to
calculate differential distributions discussed in the following section. We
observed that both sets of gluons (linear and nonlinear) give comparable results
for the phenomenological observables. This is consistent with the expectation
that saturation plays a limited role in central-forward dijet production in
$p-p$ collisions. Therefore, given that the nonlinear KS gluon distribution comes from a
better fit to $F_2$ than its linearized version~\cite{Kutak:2012rf}, in the
following, we present only the results obtained with the nonlinear gluon
density.

The new gluons presented in this section are available publicly from the
recent version of the {\tt KS package} and can be downloaded from 
\url{http://nz42.ifj.edu.pl/~sapeta/KSgluon-2.0.tar.gz}.

%-----------------------------------------------------------------------------
\section{Differential distributions}

\begin{figure}[t]
  \begin{center}
    \includegraphics[width=0.49\textwidth]{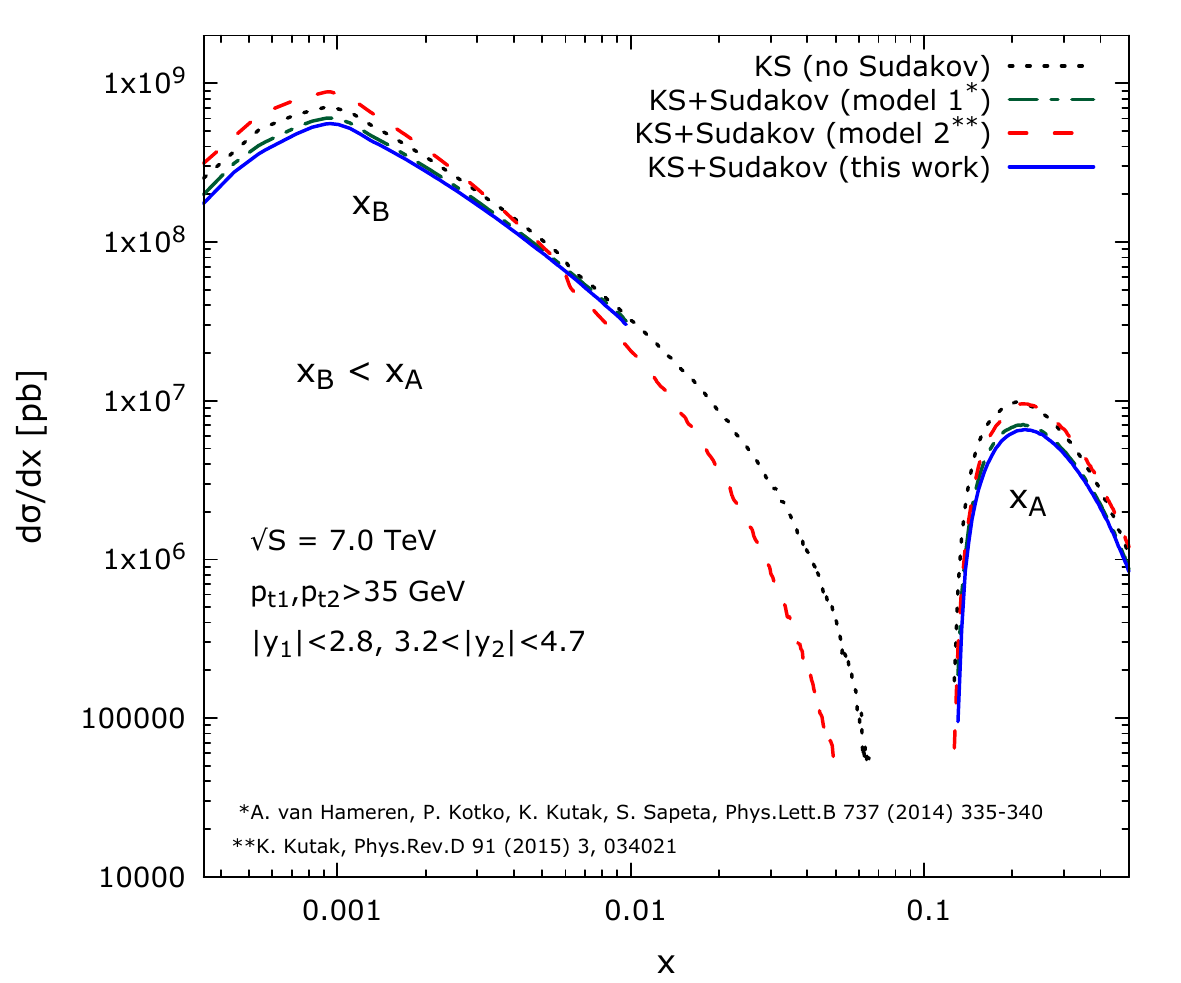}
  \end{center}
  \caption{
  Distributions of the longitudinal momentum fractions, $x_A$, $x_B$, defined in
  Eq.~(\ref{eq:xAxB}) from calculations with various version of the KS gluon
  distribution discussed in the article.
  }
  \label{fig:xdist}
\end{figure}

We now turn to the discussion of differential distributions in jets' transverse
momenta calculated in the framework described in the preceding sections.
We calculated the cross sections using the selection criteria of
CMS~\cite{Chatrchyan:2012gwa}.  The two leading jets were required to satisfy
the cuts $p_{1\perp}, p_{2\perp} > 35\,$ GeV and  $|y_1| < 2.8$,
$3.2<|y_2|<4.7$. We used the CTEQ18 NLO PDF set~\cite{Hou:2019efy} and LHAPDF
\cite{Buckley:2014ana} for the collinear PDFs and the KS gluon distributions  with and without
Sudakov for the \gluonTMDs.

Our calculations have been performed and cross checked using two independent
Monte Carlo programs \cite{vanHameren:2016kkz,Kotko_LxJet} implementing the high
energy factorization together with the off-shell matrix element calculated
following the methods of
Refs.~\cite{vanHameren:2012uj,vanHameren:2012if,Kotko:2014aba}.  We used the
average transverse momentum of jets as both the renormalization and
factorization hard scales.

\begin{figure}[t]
  \begin{center}
    \includegraphics[width=0.49\textwidth]{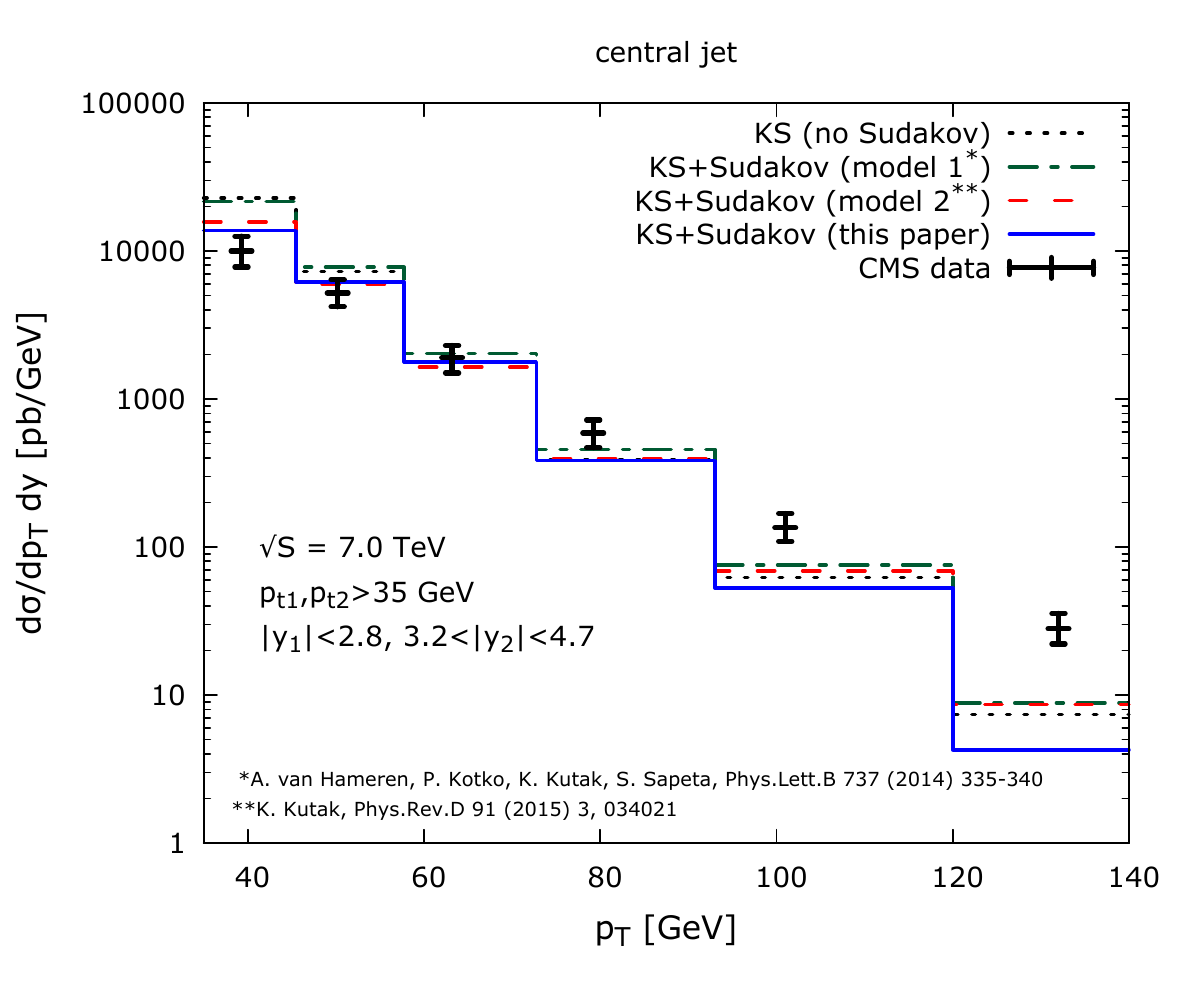}
    \includegraphics[width=0.49\textwidth]{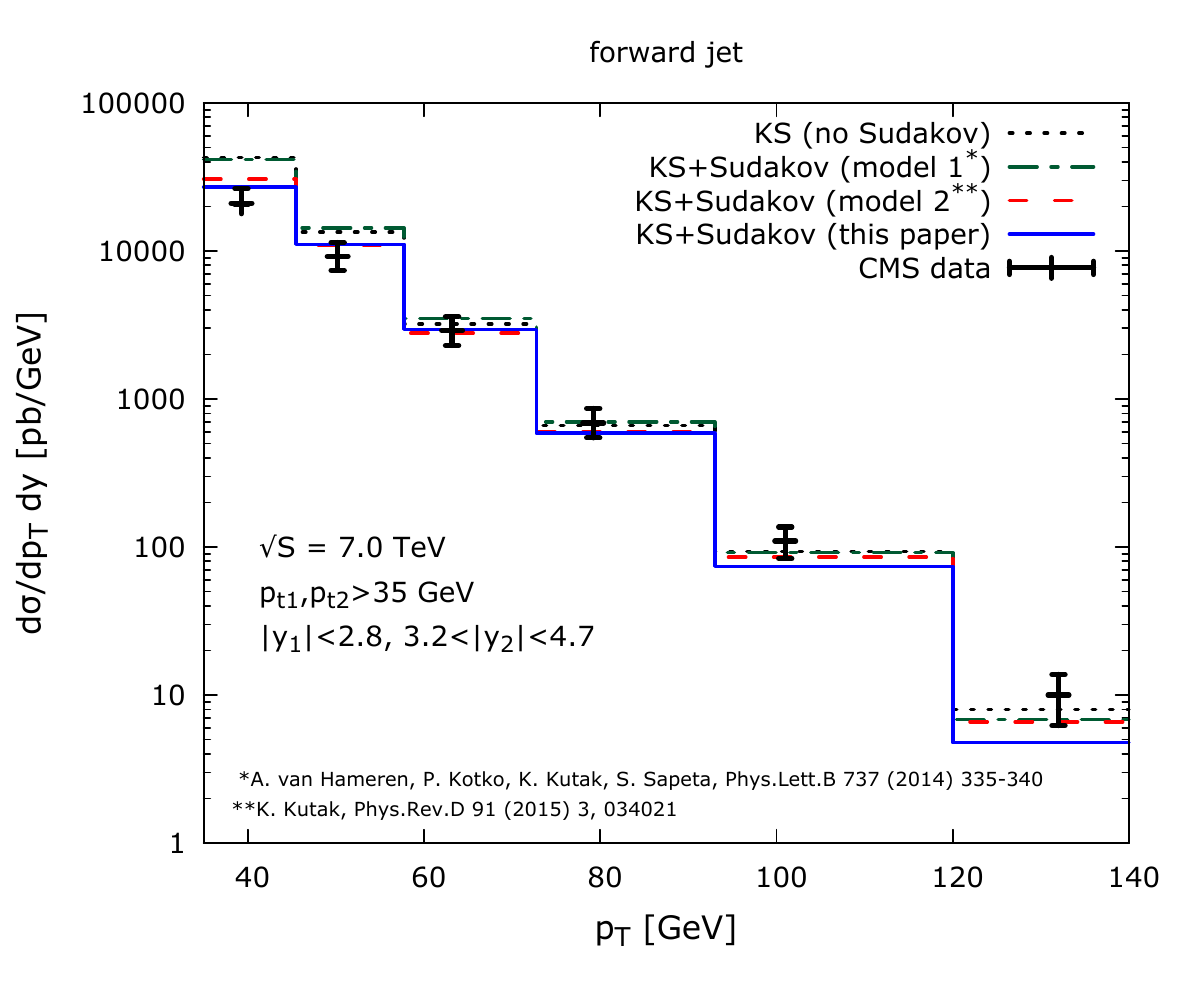}
  \end{center}
  \caption{
  The transverse momentum spectra of the central (left) and the forward (right)
  jets obtained with the KS gluon distribution, with and without Sudakov
  effects, computed for the central value of the factorization and
  renormalization scale, compared to CMS data~\cite{Chatrchyan:2012gwa}.
  }
  \label{fig:ptdist1}
\end{figure}

\begin{figure}[t]
  \begin{center}
    \includegraphics[width=0.49\textwidth]{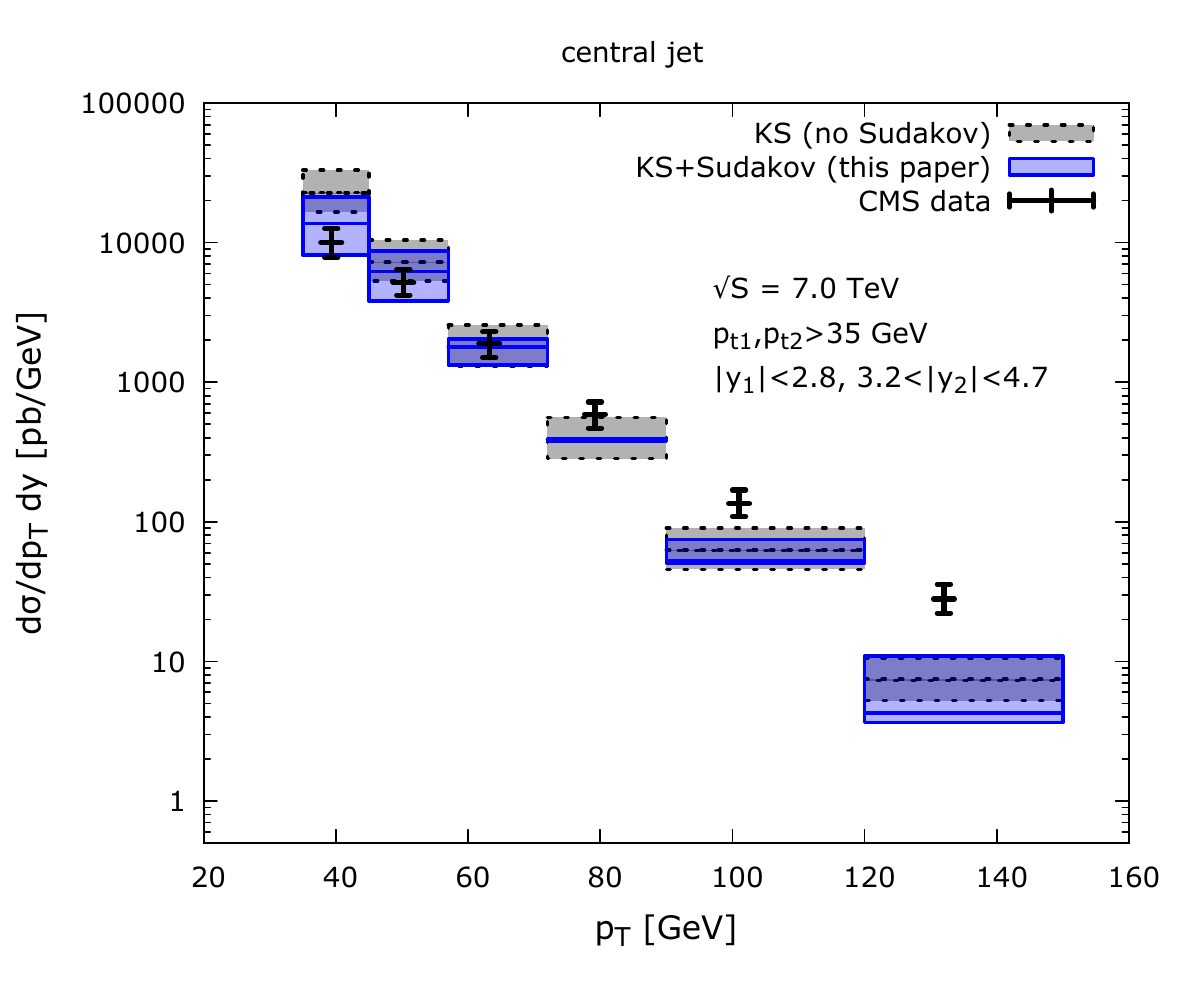}
    \includegraphics[width=0.49\textwidth]{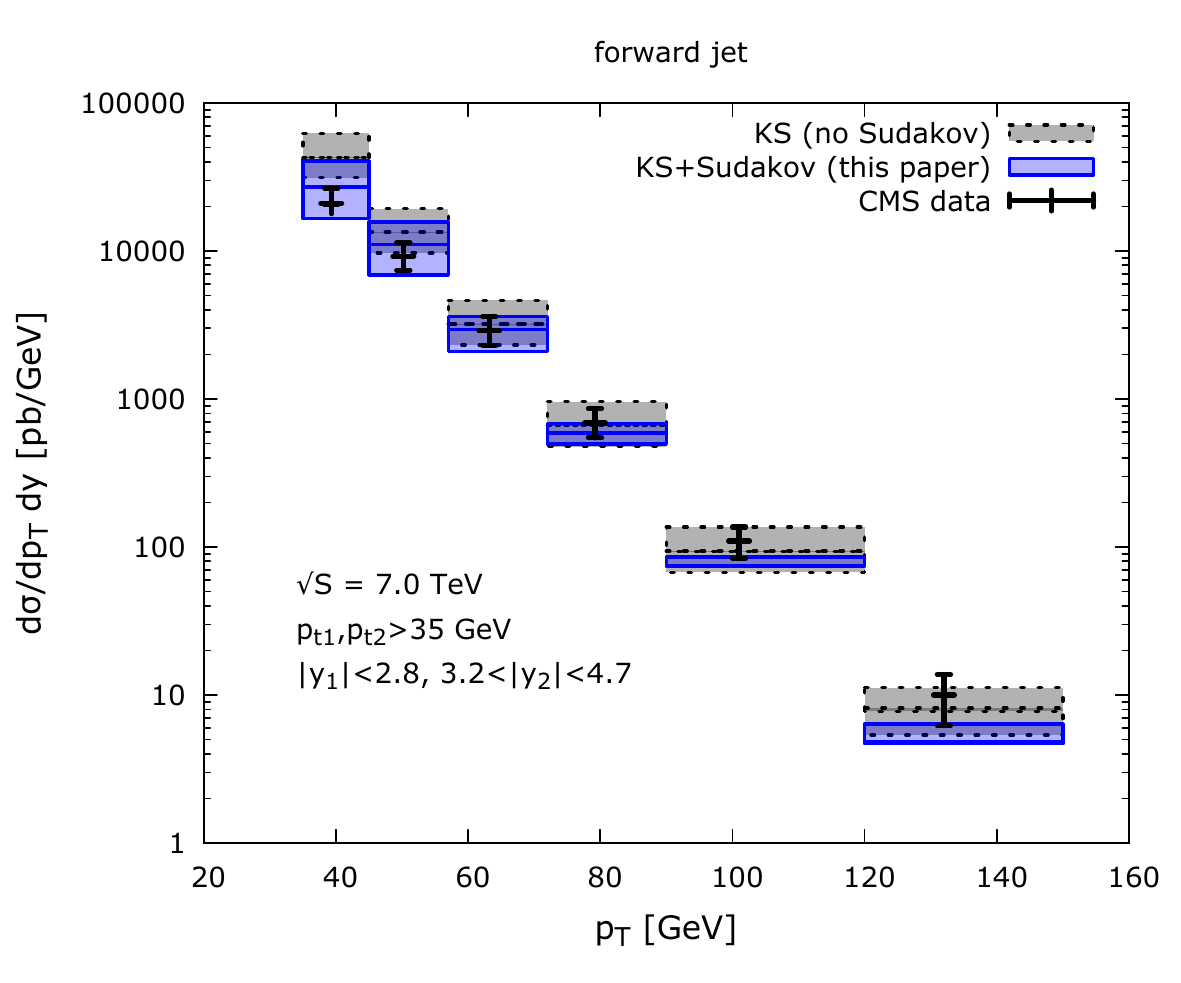}
  \end{center}
  \caption{
  As Fig.~\ref{fig:ptdist1} but we only show predictions obtained with the
  original KS gluon distribution and the predictions with the KS gluon
  distribution with Sudakov from this
  work. The bands correspond to varying the renormalization and factorization
  scale by factors $2^{\pm 1}$.
  }
  \label{fig:ptdist2}
\end{figure}

\begin{figure}[t]
  \begin{center}
    \includegraphics[width=0.49\textwidth]{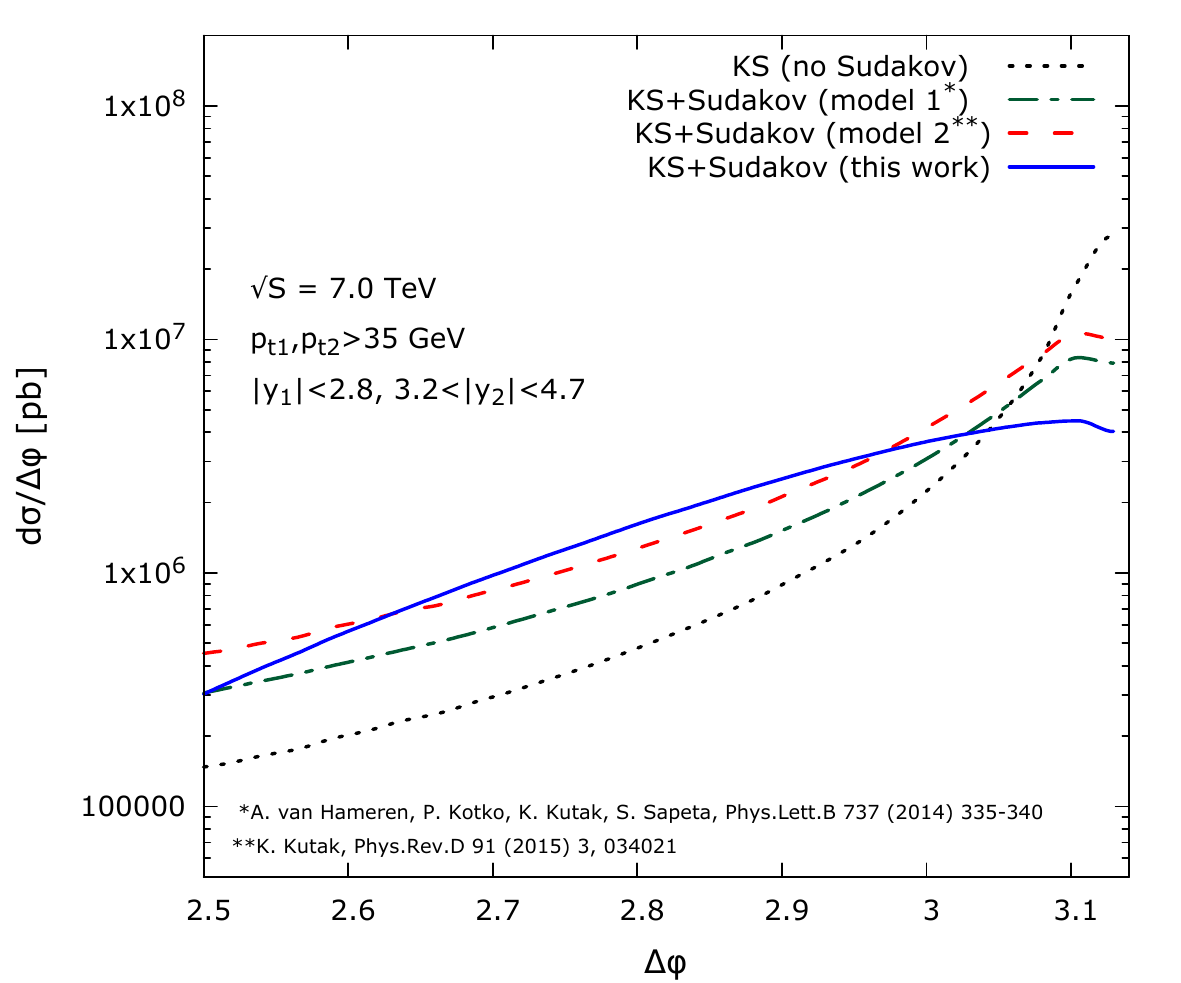}
    \includegraphics[width=0.49\textwidth]{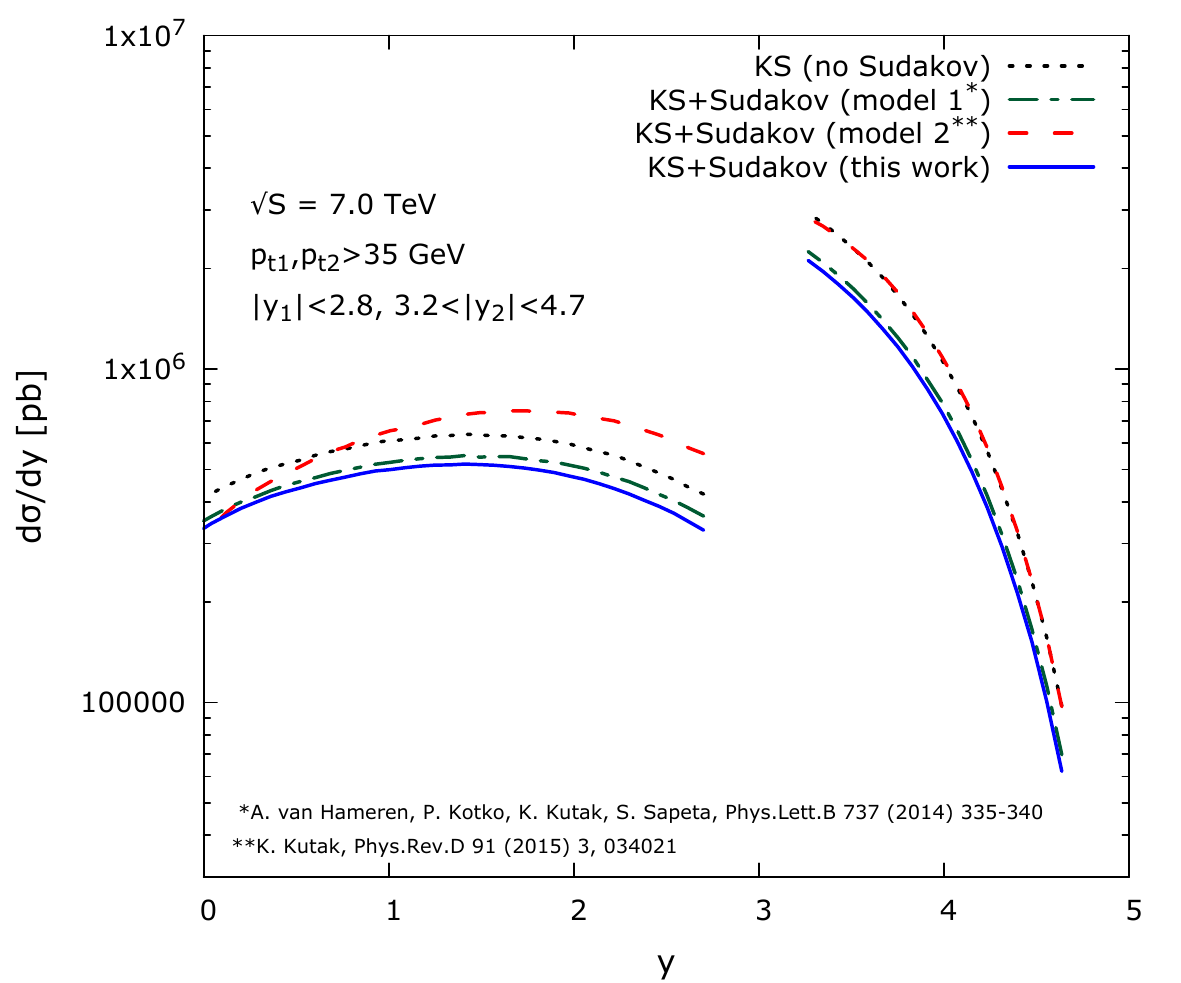}
  \end{center}
  \caption{
  Differential cross sections as functions of the azimuthal distance between the
  jets $\Delta\phi$~(left) and jet rapidities (right) obtained with the KS gluon
  distribution with and without Sudakov effects.
  }
  \label{fig:dphi}
\end{figure}

We start by showing in Fig.~\ref{fig:xdist} distributions of the longitudinal
momentum fractions probed by the central-forward dijet configurations. These
results are consistent with the discussion of Section~\ref{sec:introduction}, in
particular Eq.~(\ref{eq:xAxB}),  and provide justification to the use of the
hybrid factorization formula~(\ref{eq:hef}).

If Fig.~\ref{fig:ptdist1} we show differential cross sections as function of
the momenta of the forward and central jets. We compare central values of
various predictions  which differ by the \gluonTMDs\ used in the HEF
formula~(\ref{eq:hef}). The black dotted histograms correspond to the gluon
without Sudakov, while the other three histograms use gluons with some form of
Sudakov resummation. The main result of this paper is shown as a blue solid
line, while the green and the red dashed curves correspond to the naive Sudakov
modelling of Refs.~\cite{vanHameren:2014ala, Kutak:2014wga}.

We see that the predictions from this work, with the Sudakov effects included,
tend to describe the data better than the predictions without the Sudakov,
especially in the region of small~$p_{\perp}$. However, the overall effect of
the Sudakov form factor is not very strong for this particular observable.

If Fig.~\ref{fig:ptdist2}, we show the same distributions of transverse momenta,
but, here,  we plot only two models (without Sudakov and with Sudakov from
Section~\ref{sec:sudakov}). This time, we show also the theoretical errors,
estimated by the usual renormalization and factorization scale variation by
the factors $2^{\pm 1}$.

We observe good agreement of our predictions with the CMS
data~\cite{Chatrchyan:2012gwa}, except the tail of the central-jet transverse
momentum distribution. One has to remember however that, following
Eq.~(\ref{eq:xAxB}), the tails of $p_{\perp}$ distributions are sensitive to the
region of large $x$, where, in principle, the \gluonTMDs\ are not valid.
Indeed, we have seen in our calculation that the KS gluon distribution with the Sudakov can
sometimes get negative for larger $x$ values. We interpret that as a sign of
going outside of the validity region of the gluon distribution and, hence, in
such situations, we set it to zero in the cross section calculation.

In Fig.~\ref{fig:dphi} (left) we compare predictions for the distribution of the
azimuthal angle between the two leading jets (aka azimuthal decorrelations).
Again, we show results corresponding to calculations with and without the
Sudakov. We observe that inclusion of Sudakov effects leads to 
qualitatively the same modification of~$\Delta\phi$ distributions. Namely, the
region of large~$\Delta\phi$ is depopulated w.r.t.\ the result without Sudakov,
while the opposite happens in the region of smaller~$\Delta\phi$.

While qualitatively the predictions from KS gluon distribution + Sudakov from this work look
similar to the earlier Sudakov models, quantitatively those cross sections
differ to a certain degree, as seen in Fig.~\ref{fig:dphi}.  In particular,
the models 1 and 2, lead to convex functions for the azimuthal decorrelations,
while the Sudakov of this study produces a concave curve.

We would also like to mention that the predictions using model 1 were shown to
successfully reproduce the shapes of preliminary CMS data for the azimuthal
decorrelations~\cite{vanHameren:2014ala}. Since, as of today, these data are not
published, we refrain from comparing them with the predictions of this work. We
would only like to comment that, based on the comparison shown in
Fig.~\ref{fig:dphi}, we expect the predictions from this study to be largely
compatible with the earlier naive models, within theoretical errors.

Finally, in Fig.~\ref{fig:dphi} (right) we show rapidity distributions resulting
from the various versions of the KS gluon distribution, for the central
and the forward jet. We see marked differences between predictions without and
with Sudakov. Interestingly, inclusion of the Sudakov from this work suppresses
both the central and the forward jet distribution, and this is largely
consistent with the naive model 1. However, model 2 shows enhancement (central
jet) or almost no effect (forward jet) in the rapidity differential cross
sections.

The results presented in this section took advantage of the recent developments
in the merging of the small-$x$ dynamics and the resummation of the Sudakov
logarithms. Such calculations were not available at the time of our previous
study~\cite{vanHameren:2014ala}, thus we had to resort to simple models of the
Sudakov resummation. The calculations presented in this work are much more sound
from the theory point of view.  The new results show similar (or better) quality
in the description of the transverse momentum spectra as compared to the methods
of Refs.~\cite{vanHameren:2014ala,Kutak:2014wga}.  Likewise, in this work, we
obtain predictions for azimuthal decorrelations, which are much more sensitive
to the Sudakov resummation procedure. In particular, we see that the present
approach gives a somewhat stronger suppression of the correlation peak.

Even though the predictions from this study are close to those from our earlier
calculation, their theoretical status is much higher since, in this work, we
used the proper Sudakov factor derived from first principles in QCD. And this
was the main motivation behind the study presented in this paper.

We believe that our upgraded theoretical setup is useful in particular for the
observables like the $\Delta\phi$ distribution, where the Sudakov effects are
strong, as shown in Fig.~\ref{fig:dphi}. 
The central-forward dijet production process provided us an excellent ground for
validation of our framework. The latter can be used now to study small-$x$
dynamics, in particular the saturation effects, which are more pronounced in
the production of the forward-forward dijet system, in particular in proton-lead
collisions.

%-----------------------------------------------------------------------------
\section{Summary}

We discussed Sudakov effects in central-forward dijet
production at LHC energies within the framework of high energy
factorization.
Our study was triggered by recent progress on consistent merging of
Sudakov resummation with the small-$x$ effects, which allowed us to compute
hard-scale dependent \gluonTMDs.
As explained in Section~\ref{sec:sudakov}, we were able to combine the
phenomenologically successful KS gluon distribution~\cite{Kutak:2012rf} with the Sudakov
factors directly in momentum space.

In our study, we used the Sudakov factors derived within perturbative QCD in
Refs.~\cite{Mueller:2013wwa, Mueller:2012uf, Sun:2014gfa, Sun:2015doa,
Mueller:2016xoc}. For comparison, we also used simpler Sudakov models employed
in our earlier studies~\cite{vanHameren:2014ala, Kutak:2014wga}.

We have calculated theoretical predictions for the differential cross sections
as functions of~$p_{\perp}$ of the central and the forward jet, as well
as azimuthal distance between the jets.
The results are largely consistent with our earlier predictions based on simple
phenomenological Sudakov models. We also achieved good description of CMS data
for $p_{\perp}$ distributions. Finally, we presented predictions for dijet
azimuthal decorrelations. 

It is worth emphasising that our framework is relatively simple and all the
parametrizations of non-perturbative physics were taken from external analyses,
as explained in Section~\ref{sec:sudakov}.  Hence, no additional parameters were
introduced in the calculation of the results presented in this work.

Overall, we conclude that the Sudakov resummation has a moderate effect on
$p_{\perp}$ spectra and a fairly sizable effect on the shapes of decorrelations.
This is consistent with earlier phenomenological
studies~\cite{vanHameren:2014ala, vanHameren:2019ysa}, which showed preference
for gluons with Sudakov effects included.

Our future work will concern developing a full set of \TMD\ gluon distributions exhibiting saturation effects and the Sudakov resummation, following the same perturbative calculations we used in the present paper. Such \TMDs\ are necessary to confirm our previous calculations for forward-forward dijets \cite{vanHameren:2019ysa} that show interplay of saturation effects and Sudakov effects consistent with the ATLAS data, where, however, the more naive Sudakov model was used. 

Furthermore, in the future, we plan to address the dijet production in DIS, and
a good understanding of the interplay of Sudakov effects and saturation is
needed in order to provide robust predictions for the EIC \cite{Accardi:2012qut}
jet observables. We expect that by starting with central rapidities and going to
more forward rapidities, one will be able to incrementally see the increasing
importance of saturation effects and disentangle them from Sudakov effects.

%-----------------------------------------------------------------------------
\section*{Acknowledgements}
We would like to thank Bo-Wen Xiao for useful explanations of some details of
Ref.~\cite{Stasto:2018rci}.  Piotr Kotko is partially supported by the Polish
National Science Centre grant no.\ 2018/31/D/ST2/02731.  Andreas van Hameren is
partially supported by the Polish National Science Centre grant no.\
2019/35/B/ST2/03531.  Sebastian Sapeta is partially supported by the Polish
National Science Centre grant no.  2017/27/B/ST2/02004. This work has received
funding from the European Union’s Horizon 2020 research and innovation
programme under grant agreement No. 824093.

%-----------------------------------------------------------------------------
\bibliographystyle{unsrt} % sorts in the order of apperance
\bibliography{sudakov}

\end{document}